\documentclass[aps,nofootinbib,prl,reprint]{revtex4-1}

\pdfoutput=1
\usepackage{graphicx}

\usepackage{amsmath,amssymb,mathrsfs}

\usepackage[bookmarks=false]{hyperref} 
\hypersetup{pdfstartview=FitH,pdfhighlight=/O,colorlinks=false}

\newcommand{\p}{\partial}

\begin{document}

\title{Entropy of Non-Extremal Black Holes from Loop Gravity}
\author{Eugenio~Bianchi\\[-.5em] ${}$} 
\affiliation{Perimeter Institute for Theoretical Physics\\
31 Caroline St.N., Waterloo ON, N2J 2Y5, Canada\\}
\date{April 22, 2012}

\begin{abstract}
We compute the entropy of non-extremal black holes using the quantum dynamics of Loop Gravity. The horizon entropy is finite, scales linearly with the area $A$, and reproduces the Bekenstein-Hawking expression $S = A/{4G\hbar}$ with the one-fourth coefficient for all values of the Immirzi parameter. The near-horizon geometry of a non-extremal black hole -- as seen by a stationary observer -- is described by a Rindler horizon. We introduce the notion of a quantum Rindler horizon in the framework of Loop Gravity. The system is described by a quantum surface and the dynamics is generated by the boost Hamiltonion of Lorentzian Spinfoams. We show that the expectation value of the boost Hamiltonian reproduces the local horizon energy of Frodden, Ghosh and Perez. We study the coupling of the geometry of the quantum horizon to a two-level system and show that it thermalizes to the local Unruh temperature. The derived values of the energy and the temperature allow one to compute the thermodynamic entropy of the quantum horizon. The relation with the Spinfoam partition function is discussed.
\end{abstract}


\maketitle
There is strong theoretical evidence that Black Holes have a finite thermodynamic entropy equal to one quarter the area $A$ of the horizon \cite{Bekenstein:1973ur}. Providing a microscopic derivation of the Bekenstein-Hawking entropy\footnote{Units as follows: we measure time intervals in meters and temperatures in Joules, so that the speed of light and Boltzman's constant are set to one,  $c=1$ and $k_B=1$. We keep explicit the dependence on Newton's constant $G$ and Planck's constant $\hbar$.}
\begin{equation}
S_{\text{BH}}=\frac{A}{4\: G \hbar}\label{eq:S-BH}
\end{equation}
is a major task for a candidate theory of quantum gravity. Loop Gravity \cite{Rovelli:2004tv} has been shown to provide a geometric explanation of the finiteness of the entropy and of the proportionality to the area of the horizon \cite{Rovelli:1996dv}. The microstates are quantum geometries of the horizon \cite{Bianchi:2010qd}. What has been missing until recently is the identification of the near-horizon quantum dynamics and a derivation of the universal form of the  Bekenstein-Hawking entropy with its $1/4$ prefactor. This is achieved in this letter.\\

We consider non-extremal Black Holes of the Kerr-Newman family. Their near-horizon geometry -- as seen by a stationary observer -- is described by a Rindler horizon. Frodden, Ghosh and Perez have identified a local notion of near-horizon energy \cite{Frodden:2011eb}. We show that in Loop Gravity there is a natural quantum version of horizon states and of horizon energy: (i) the states live in a tensor product of $SU(2)$ representation spaces and (ii) the horizon energy is given by the boost Hamiltonian acting on the $\gamma$-simple unitary representations of the Lorentz group, where $\gamma$ is the Immirzi parameter. These representations are exactly the ones that appear in the covariant dynamics of Loop Gravity and define the Lorentzian Spinfoam path integral \cite{Engle:2007uq}. 

We compute the energy and the temperature of the quantum horizon. The expectation value of the energy reproduces the Frodden-Ghosh-Perez energy,
\begin{equation}
E=\frac{A}{8\pi G}\,l^{-1}\;,
\end{equation}
where $l$ is the distance of the stationary observer from the horizon and is assumed to be large compared to the Planck scale $l\gg \sqrt{G\hbar}$. The horizon temperature, computed at the leading order in the acceleration $a=l^{-1}$, reproduces the Unruh temperature \cite{Unruh:1976db}
\begin{equation}
T=\frac{\hbar a}{2\pi}\;.
\end{equation}
Using the Clausius relation these two equations give immediately the Bekenstein-Hawking entropy with coefficient $1/4$. The Immirzi parameter enters only in quantum gravity corrections to the entropy.

The notion of quantum horizon rests on a defining property of the Lorentzian Spinfoam dynamics, the $\gamma$-simple unitary representations. In particular the derivation of the thermodynamic properties of the quantum horizon does not require the identification of a semiclassical regime away from it. We discuss also the relation with the Spinfoam partition function, its semiclassical limit, and the relation with the Gibbons-Hawking path-integral derivation of the entropy \cite{Gibbons:1976ue}.

\vspace{-1em}
\section{Near-horizon geometry}
\vspace{-1em}

Consider a Schwarzschild Black Hole of mass $M$, and a stationary observer at proper distance $l$ from the horizon. We assume the observer to be close to the horizon, $l\ll r_S=2GM$. The near-horizon metric in coordinates $\{t,l\}$ is of the Rindler form
\begin{equation}
ds^2=-(\kappa l)^2dt^2+dl^2+r_S^2d\Omega^2\;, \label{eq:Rindler}
\end{equation}
where  $\kappa= \frac{1}{2r_S}$ is the surface gravity and $d\Omega^2$ the measure on the unit sphere. The four-velocity of such an observer is $u^\mu=\frac{1}{\kappa l} (\frac{\p}{\p t})^\mu$ and her four-acceleration is $a^\mu= u^\nu\nabla_\nu u^\mu=a\, (\frac{\p}{\p l})^\mu$, with acceleration $a=l^{-1}$. Notice that the acceleration depends only on the distance from the horizon, and in particular it is independent from the mass of the Black Hole. The surface $l=0$ is a non-degenerate Killing horizon, the Rindler horizon. The analysis above generalizes to non-extremal Black Holes of the Kerr-Newman family \cite{Wald:1984rg} and to non-extremal isolated horizons \cite{Ashtekar:2000sz}. In the first case the stationary observers follow the integral curves of the Killing vector field $\chi=\frac{\p}{\p t}+\Omega \frac{\p}{\p \phi}$, where $\frac{\p}{\p t}$ and $\frac{\p}{\p \phi}$ are the Killing fields associated with the stationarity and axisymmetry of the geometry, and $\Omega$ is the horizon angular momentum. These are the unique stationary observers that coincide with the locally non-rotating observers of \cite{Wald:1984rg} or ZAMOs of \cite{Thorne:1986iy} as $l\to0$. Their acceleration is again $a=l^{-1}$, where $l$ is the proper distance from the horizon.

There is a natural notion of near-horizon energy associated with the stationary observers discussed above, it is the Frodden-Ghosh-Perez energy \cite{Frodden:2011eb}
\begin{equation*}
E=\frac{1}{8\pi G}\int \nabla^\mu u^\nu\;dS_{\mu\nu}  \; =\;\frac{1}{8\pi G}\int_{S}\sqrt{h}\;n_\mu a^\mu \; d^2x\;,
\end{equation*}
where $dS_{\mu\nu}=(u_\mu n_\nu-u_\nu n_\mu) \sqrt{h} \,d^2\sigma$ is the area element on the surface, $n^\mu=(\frac{\p}{\p l})^\mu$ is the space-like normal to the horizon and to the 4-velocity of the stationary observer, $S$ is a stationary surface at proper distance $l$ from the horizon, and $\sqrt{h}$ its area-density. As $a^\mu n_\mu=a$ is constant on $S$, the near-horizon energy simply evaluates to
\begin{equation}
E=\frac{A}{8\pi G}a\;,  \label{eq:E=Aa}
\end{equation}
where $A$ is the area of the horizon. A physical arguments for this expression for the energy is the following: If we drop a test particle of mass $m$ and charge $q$ from infinity to the horizon of a Kerr-Newman black hole, a stationary observer who keeps her distance $l$ from the horizon fixed will measure a local energy $\epsilon_{\text{loc}}$ of the particle. In  \cite{Frodden:2011eb} it is shown that this energy can be expressed in terms of geometric quantities only and is given by $\epsilon_{\text{loc}}=\delta E=\frac{\delta A}{8\pi G}a$, where  $\delta A$ is the variation of area of the horizon after the particle has been absorbed by the black hole.

\vspace{-1em}
\section{The quantum Rindler horizon} 
\vspace{-1em}

In Loop Gravity $SU(2)$ spin-network states represent the geometry of a spatial section of space-time having given normal $t^\mu$. The group $SU(2)$ is the little group of the local Lorentz group $SL(2,C)$ that preserves the time-like vector $t^\mu$. The covariant Spinfoam dynamics is defined by introducing $SL(2,C)$ spin-networks and requiring local Lorentz invariance. A key object in the definition of Spinfoams is a map $Y_\gamma$ that provides an injection of the $SU(2)$ representation $V^{(j)}$ of spin $j$ into the $SL(2,C)$ unitary representation $\mathcal{V}^{(p,k)}$ \cite{Ruhl}. The real number $p$ and the half-integer $k$ label the values of the two Casimir operators, $\vec{K}^2-\vec{L}^2=p^2-k^2+1$ and $\vec{K}\cdot\vec{L}=pk$, where $\vec{L}$ is the generator of rotations and $\vec{K}$ the generator of boosts. We call $|(p,k);j,m\rangle$ a basis of eigenstates of $\vec{L}^2$ and $L_z=\vec{L}\cdot\vec{n}$. The map $Y_\gamma$ injects $V^{(j)}$ in the lowest-weight block of $\mathcal{V}^{(p,k)}$ with the following choice of $p$ and $k$:
\begin{align*}
Y_{\gamma}: & \;\;V^{(j)}\to\;\; \mathcal{V}^{(\gamma(j+1), j)}\\
&|j,m\rangle\mapsto |(\gamma(j+1),j); j,m\rangle\;.
\end{align*}
The representation $p=\gamma(j+1)$ and $k=j$ is called $\gamma$-simple and satisfies the condition
\begin{equation}
\vec{K}=\gamma\vec{L} \label{eq:K=gL}
\end{equation}
as matrix elements on the image of the $Y_{\gamma}$. Here $\gamma$ is the Immirzi parameter and the condition (\ref{eq:K=gL}) is the one that unfreezes gravitational degrees of freedom from a topological field theory, and defines quantum general relativity in the Spinfoam approach \cite{Rovelli:2011eq}.

Consider a surface $S$ tessellated by $N$ facets $f$, each one corresponding to a link of a spin-network graph puncturing the surface. The state describing a quantum surface of given area and with normal in direction $\vec{n}$ has the form $|s\rangle=\bigotimes_f |j_f\rangle$, where $|j_f\rangle$ is the state with maximum magnetic number in direction $\vec{n}$
\begin{equation}
|j\rangle\equiv|(\gamma(j+1),j); j,+j\rangle\;.\nonumber
\end{equation}
The area operator is $A=8\pi G\hbar \gamma\sum_f |\vec{L}_f|$, so that the state $|s\rangle$ has area $A=\sum_f A_f$ with
\begin{equation}
A_f=8\pi G\hbar \gamma j_f\;.\label{eq:area}
\end{equation}

Consider the operator $H$ defined as
\begin{equation}
H=\sum_f \hbar\, K^z_f\, a\;,
\end{equation}
where $a$ is a positive real number and $K_z=\vec{K}\cdot\vec{n}$. This operator generates evolution along a uniformly accelerated trajectory. Therefore it defines evolution in the presence of a quantum Rindler horizon, as seen from an observer with acceleration $a$, and represents the energy of the system in the accelerated frame.

The quantum Rindler horizon is defined by the evolution of the state $|s\rangle$ with the dynamics $H$,
\begin{equation}
|s_t\rangle=U(t)\;|s\rangle \nonumber
\end{equation}
where $U(t)$ is the unitary operator representing a Lorentz boost with rapidity $a t$,
\begin{equation}
U(t)=e^{\frac{i}{\hbar}H t}=\otimes_{f} \exp(i K^z_f \,a t)\;. \nonumber
\end{equation}

\vspace{-1em}
\section{Energy of the quantum horizon} 
\vspace{-1em}

Now we show that the operator $H$ measures the energy of the quantum horizon. The expectation value of $H$ on the quantum horizon state $|s\rangle=\bigotimes_f |j_f\rangle$ can be easily computed using the following result: on the image of $Y_\gamma$ the matrix elements of the boost generator simply evaluate to
\begin{equation}
 \langle(\gamma(j+1),j); j,m'|K_z|(\gamma(j+1),j); j,m\rangle=\gamma\, m\;\delta_{mm'}\;. \nonumber
\end{equation}
As a result, we find
\begin{equation}
E\equiv\langle s | H |s\rangle=\sum_f \hbar\, \gamma j_f\,a\;=\frac{\sum_f A_f}{8\pi G} a\;,\label{eq:E}
\end{equation}
where in the third equality we have used the expression of the area (\ref{eq:area}). The expectation values of the boost Hamiltonian $H$ reproduce the expression (\ref{eq:E=Aa}) of the classical near-horizon energy. This result shows that the Hamiltonian that generates the quantum Rindler horizon also measures its energy.\\

Eigenstates of the energy $|E\rangle=\bigotimes_f |\lambda_f\rangle$ are labeled by continuous parameters $\lambda_f$, where we define
\begin{equation}
|\lambda\rangle\equiv|(\gamma(j+1),j); \lambda,+j\rangle\;\nonumber
\end{equation}
as the simultaneous eigenstate of $K_z=\lambda$ and $L_z=+j$  \cite{Ruhl}. The energy of the state $|E\rangle$  is $E=\sum_f\hbar \lambda_f a$.

\vspace{-1em}
\section{Temperature of the quantum horizon}
\vspace{-1em}

The energy levels $|E\rangle$ of the quantum horizon are highly degenerate. Moreover the energy does not commute with the area of the quantum horizon. Now we show that a consequence of this fact is that the quantum horizon has a finite temperature $T$. 

To measure the temperature of the quantum horizon we introduce a thermometer: a two-level detector coupled to the horizon \cite{Unruh:1976db,DeWitt:1979}. Let $|0\rangle$ be its ground state, $|1\rangle$ the excited state, and $H_D=\epsilon_0 |0\rangle\langle 0|+\epsilon_1 |1\rangle\langle 1|$ the Hamiltonian. The energy level separation is $\Delta \epsilon=\epsilon_1-\epsilon_0$. The detector interacts with the horizon via a potential
\begin{equation}
V= g \;  (|0\rangle\langle1| + |1\rangle\langle0|)\;\psi\;\phi\;,
\end{equation}
where $g$ is a coupling constant, $\psi=|E_0\rangle\langle E_1| + |E_1\rangle\langle E_0|$ couples different energy levels of the horizon, and $\phi$ is the hermitian operator that applied on the trivial representation $|\Omega\rangle$ of the Lorentz group -- a facet in the \emph{vacuum state} -- gives a state of definite area $|j\rangle$,
\begin{equation}
\phi\,  |\Omega\rangle=|j\rangle\;.
\end{equation}
The transition rate $\Gamma_+$  for the two-level system to be excited from its ground state, and $\Gamma_-$ for the decay, can be easily computed using Fermi's golden rule
\begin{equation}
\hspace{-.5em}\Gamma_{\pm}=\frac{g^2}{\hbar^2}\int_{-\infty}^{+\infty}\hspace{-1.5em}d\tau\;\;e^{-\frac{i}{\hbar}(E_1-E_0\mp \Delta\epsilon)\tau}\;\langle\Omega|\, \phi \,e^{iK_z a \tau}\,\phi\,|\Omega\rangle\;.  \label{eq:Gamma+}
\end{equation}
For $\hbar a\ll \Delta \epsilon\ll E_1-E_0$, we find that the equilibrium population of the two-level system has ratio $R$ that is independent of the $j$, $\gamma$, $E_0$, $E_1$ and given by
\begin{equation}
R=\frac{\Gamma_+}{\Gamma_-}\approx \exp({-\frac{2\pi}{\hbar a}\Delta\epsilon})\;.\label{eq:R}
\end{equation}
The result is a Boltzman distribution of temperature $T$,
\begin{equation}
T=\frac{\hbar a}{2\pi}\;.\label{eq:T}
\end{equation}
The temperature of the quantum Rindler horizon measured by an observer with acceleration $a$ coincides with the Unruh temperature.\\

We give a derivation of this result. Consider the transition amplitute $\mathcal{A}_+$ from the initial state $|\text{in}\rangle$
\begin{equation}
|\text{in}\rangle=|E_1,\Omega\rangle|0\rangle\nonumber
\end{equation}
with the detector in the ground state and the horizon having energy $E_1$ and one facet in its vacuum state $|\Omega\rangle$, to a final state $|\text{out}\rangle$
\begin{equation}
|\text{out}\rangle=|E_0,\lambda\rangle|1\rangle\nonumber
\end{equation}
with the horizon being in an eigenstate with energy $E_0+\hbar\lambda a$ and the detector excited. To first order in perturbation theory,
\begin{align*}
\mathcal{A}_+=&\, -\frac{i}{\hbar}\int_{-\infty}^t\hspace{-1em}dt'\;\langle\text{out}|V_I(t')|\text{in}\rangle=\\
=&\; -\frac{i}{\hbar}\,g\int_{-\infty}^t\hspace{-1em}dt'\;e^{-\frac{i}{\hbar}(E_1-E_0+\Delta\epsilon)t'}\langle\lambda |e^{i K_z a t'}\phi|\Omega\rangle
\end{align*}
where $V_I(t)=e^{\frac{i}{\hbar}H t}V e^{-\frac{i}{\hbar} H t}$ is the potential in the interaction picture, and in the second line we used the invariance property of the trivial representation $e^{-\frac{i}{\hbar} H t}|\Omega\rangle=|\Omega\rangle$. The probability $w_+$ of a transition to a final state with excited detector and state $|\lambda\rangle$ of the facet unobserved is given by the sum over final states of the modulus squared of the amplitude
\begin{align*}
w_+=&\;\int_{-\infty}^{+\infty}\hspace{-1em}d\lambda \;|\mathcal{A}_+|^2
=\; \frac{g^2}{\hbar^2} \int_{-\infty}^{t}\hspace{-1em}d t'\int_{-\infty}^{t}\hspace{-1em}d t''\;\\[.5em]
&\times e^{-\frac{i}{\hbar}(E_1-E_0+\Delta\epsilon)(t''-t')}\langle\Omega|\, \phi \,e^{iK_z a (t''-t')}\,\phi\,|\Omega\rangle
\end{align*}
where we have used the completeness of the $|\lambda\rangle$ basis to perform the integral over $\lambda$. Changing variables to $t'+t''$ and $\tau=t''-t'$, and considering the probability of transition per unit time $\Gamma_+=\dot{w}_+$, we recover the transition rate (\ref{eq:Gamma+}). The rate  $\Gamma_-=\dot{w}_-$ is easily obtained via a similar calculation.

The basis $|\lambda\rangle$ of eigenstates of the energy of the facet $f$ can be used to obtain a closed formula for the transition rate,
\begin{align}
\Gamma_{\pm}=&\;\frac{g^2}{\hbar^2}\int_{-\infty}^{+\infty}\hspace{-1.5em}d\lambda\;\int_{-\infty}^{+\infty}\hspace{-1.5em}d\tau\;\;e^{-\frac{i}{\hbar}(E_1-E_0\mp \Delta\epsilon)\tau}\;e^{i\lambda a \tau} |\langle\lambda|\,\phi\,|\Omega\rangle|^2\nonumber\\
=&\;\frac{g^2}{\hbar}\int_{-\infty}^{+\infty}\hspace{-1.5em}d\lambda\;\delta(E_1-E_0\mp \Delta\epsilon-\hbar\lambda a)|\langle\lambda |j\rangle|^2\nonumber\\
=&\;\frac{g^2}{\hbar^2 a}\;|\langle\lambda_0= \frac{E_1-E_0\mp \Delta\epsilon}{\hbar a}|j\rangle|^2\;.\label{eq:delta}
\end{align}
In the second line we have performed the integral over $\tau$ and used $|\langle\lambda|\,\phi\,|\Omega\rangle|^2=|\langle\lambda |j\rangle|^2$, and in the third integrated $\lambda$ against the Dirac delta $\delta(E_1-E_0\mp \Delta\epsilon-\hbar\lambda a)$.

The overlap coefficient between eigenstates of the boost Hamiltonian and eigenstates of the area can be computed using the methods developed in \cite{Ruhl}. 
\begin{figure}[t]
\includegraphics[width=.45\textwidth]{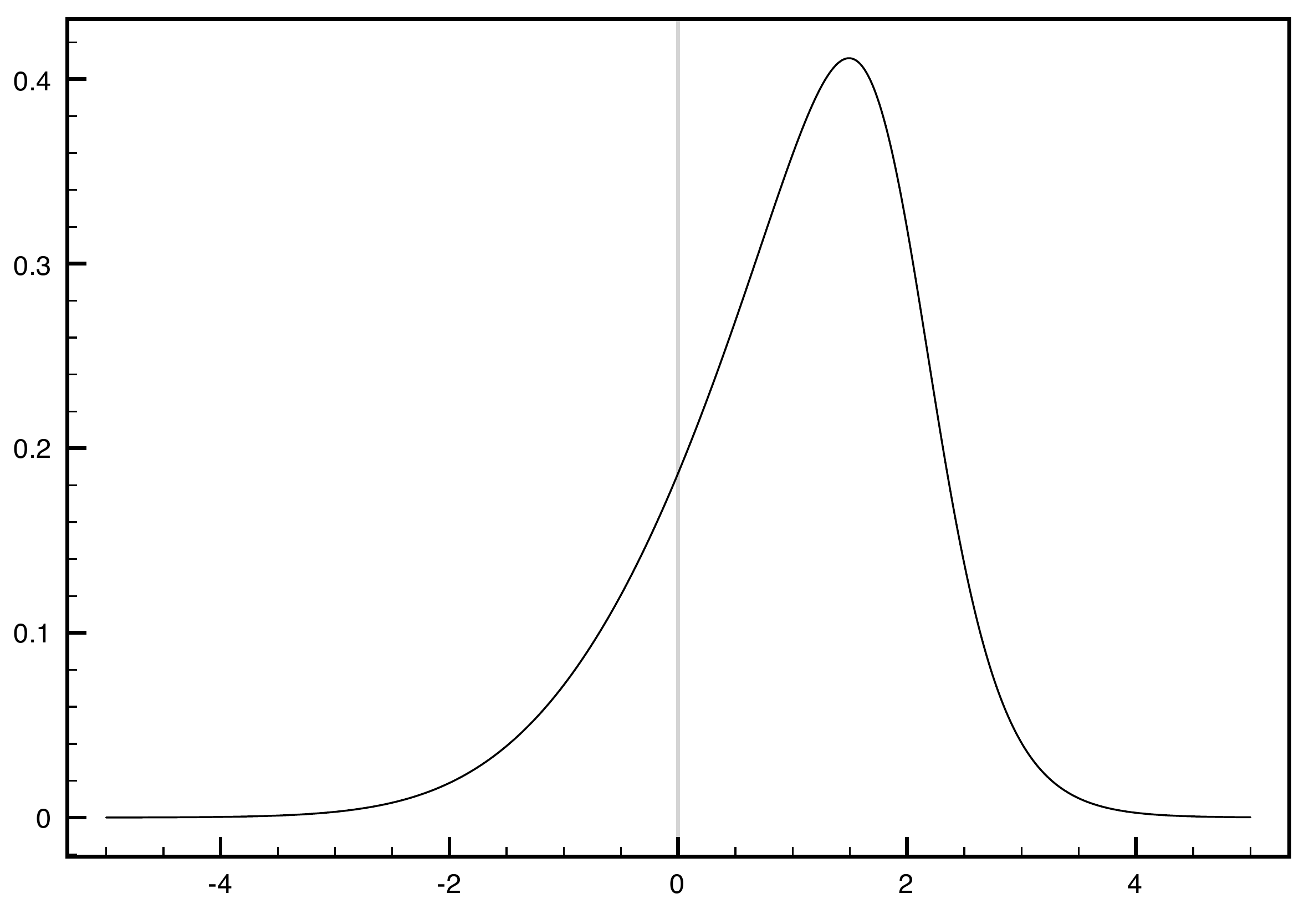}
\caption{The distribution $\varrho(\lambda)=|\langle\lambda|j\rangle|^2$ is plotted as a function of $\lambda$ for $j=1$ and $\gamma=1$. It represents the probability of finding an area eigenstate $|j\rangle$ in the state of energy $E=\hbar \lambda a$. The average is $\langle\lambda\rangle=\gamma j$, the dispersion $\Delta \lambda=\sqrt{1+\gamma^2}\sqrt{\frac{2j+1}{2j+3}}$. There is an exponential fall-off for large $\lambda$.}
\label{fig:overlap}
\end{figure}
We obtain the closed-form expression
\begin{align}
&|\langle\lambda|j\rangle|^2=\;{\textstyle\frac{1}{2}\left| \frac{(2j+1)\Gamma(-j+i\gamma(j+1))\Gamma(\frac{1}{2}+j+\frac{i}{2}(\gamma(j+1)+\lambda))}{\Gamma(1+j+i\gamma(j+1))\Gamma(\frac{1}{2}-j+\frac{i}{2}(\gamma(j+1)+\lambda))}\right.\;\times}\nonumber\\[.5em]
&\hspace{6em}\times\;{\textstyle\left.\frac{\sin\pi(j-i\gamma(j+1))}{\cos\pi(j-i\gamma(j+1))\;+\;\cos\pi(j-i\lambda)}\right|}
\end{align}
where $\Gamma(z)$ is the Euler gamma function. See the plot in figure \ref{fig:overlap}. For large $\lambda$ and using Stirling's approximation we find that it decays exponentially as
\begin{equation}
|\langle\lambda|j\rangle|^2\approx \,c\, \lambda^{2j}\; e^{- \pi \lambda}\;.
\end{equation}
The constant $c$ depends only on $j$ and $\gamma$. Under the assumption $\hbar a\ll \Delta \epsilon\ll E_1-E_0$, the transition rates are
\begin{equation*}
\Gamma_{\pm}\approx \,c\;\frac{g^2}{\hbar^2 a} \;\,{\textstyle\Big(\frac{E_1-E_0\mp \Delta\epsilon}{\hbar a}\Big)^{2j}} \;\exp(-\pi \frac{E_1-E_0\mp \Delta\epsilon}{\hbar a})\;.
\end{equation*}
The change in time of the populations $p_1$ of the excited state is $\dot{p}_1=p_0 \Gamma_+-p_1\Gamma_-$, where $p_0$ is the population of the ground state. When the system thermalizes, $\dot{p}_1=0$ and $p_1/p_0=\Gamma_+/\Gamma_-$, so that the ratio (\ref{eq:R}) is reached. In particular, from $p_0+p_1=1$ we derive the population of the excited level
\begin{equation}
p_1=\frac{1}{1+\exp{(\frac{2\pi}{\hbar a}\Delta \epsilon})}\;.
\end{equation}
Therefore we conclude that the temperature is $T=\frac{\hbar a}{2\pi}$.\\

The temperature far from the horizon -- at infinity -- can be computed via the Tolman law with the red-shift factor of the semiclassical metric away from the horizon. Equivalently, we can think of the detector as part of the quantum geometry in proximity of the horizon -- a spin $1/2$ facet -- coupled to the electromagnetic field. The effect is that it will thermalize photons at a distance $1/a$ from the horizon to the temperature $T=\frac{\hbar a}{2\pi}$. The thermal photons escape to infinity propagating in the classical background metric of a Kerr-Newman black hole, and reach infinity with a red-shifted temperature that is the Hawking temperature.

\vspace{-1em}
\section{Entropy of the quantum horizon}
\vspace{-1em}

Having computed the energy and the temperature of the quantum horizon, we can now determine its thermodynamic entropy $S$ using the Clausius relation
\begin{equation}
\delta S=\frac{\delta E}{T}\;.\label{eq:Clausius}
\end{equation}
The energy of the horizon is $E=\sum_f \hbar \gamma j_f a$  as derived in Eq.(\ref{eq:E}). In a process in which the energy changes because of a single facet $f$, we have
\begin{equation}
\delta E= \hbar \gamma j_f a\;.\nonumber
\end{equation}
The temperature of the horizon during the process has been derived above in Eq. (\ref{eq:T}). Using $\delta S_f=\delta E/T$, we obtain an entropy per facet equal to
\begin{equation}
\delta S_f=\frac{\hbar \gamma j_f a}{\hbar a/2\pi}=2\pi \gamma j_f\;.
\end{equation}
Notice that the entropy density is independent of the acceleration $a$, or equivalently from the distance from the horizon. The entropy on the quantum horizon is obtained summing over the facets the contributions $\delta S_f$ and recognizing the expression of the area (\ref{eq:area})
\begin{equation}
S=\sum_f 2\pi \gamma j_f=2\pi\frac{\sum_f A_f}{8\pi G \hbar}\;=\;\frac{A}{4G\hbar}\;.\label{eq:S}
\end{equation}
The entropy of the quantum horizon agrees with the Bekenstein-Hawking entropy with its prefactor $1/4$.\\

Notice also that the equilibrium distribution of the detector's states is governed exactly by this entropy. In fact in the process of thermalization the conservation of energy $ \Delta \epsilon = \delta E\equiv E_1-E_0-\hbar\lambda a$ is imposed by the delta function in (\ref{eq:delta}). The population ratio is
\begin{equation}
R=\frac{\Gamma_+}{\Gamma_-}\approx \exp ({-\frac{2\pi}{\hbar a}\delta E})=\exp({-\frac{\delta A}{4 G \hbar}})\;.
\end{equation}
Loop Gravity provides a realization of the scenario proposed by Massar and Parentani in \cite{Massar:1999wg}: this relation reveals that the detector is in contact with a reservoir whose statistical entropy is $S=\frac{A}{4 G \hbar}$.

\vspace{-1em}
\section{Partition function and Spinfoams}
\vspace{-1em}

\noindent A partition function of the form
\begin{equation}
Z(\beta)=\exp -\frac{1}{8\pi G \hbar}\sum_f A_f(\beta a -2\pi)\;,
\end{equation}
where $A_f=8\pi G\hbar \,\gamma j$, reproduce the values of the energy $E$ and entropy $S$ of the quantum horizon. Using standard statistical mechanics formulae, we obtain the energy and the entropy\footnote{The parameter $\beta$  has dimensions of length, so that its relation to the temperature is $\beta=\hbar/T$.}
\begin{align*}
E=&\;-\hbar\frac{\partial \log Z}{\partial \beta}=\frac{\sum_f A_f}{8\pi G}a\;,\\[.5em]
S=&\;-\beta\frac{\partial \log Z}{\partial \beta}+\log Z=\frac{\sum_f A_f}{4 G \hbar}\;.
\end{align*}
They coincide precisely with the expressions derived before. On the other hand, the partition function $Z(\beta)$ has a natural interpretation in terms of Spinfoams.

Consider Regge's discretization of General Relativity \cite{Regge:1961px}, with a triangle of area $A_f$ on the horizon and a point at distance $l=a^{-1}$ from it. This configuration defines a tetrahedron. We assume $l\gg\sqrt{A_f}$. If the point follows an accelerated motion, at a later time $\beta$ the point has evolved -- maintaining  its distance $l$ from the triangle fixed -- defining a second tetrahedron and the spacetime geometry of a $4$-simplex. The dihedral angle $\Theta$ of the $4$-simplex at the triangle $A_f$ is simply given by $\Theta=\beta/l=\beta a$. Imposing periodicity in the Euclidean time $\beta$, defines a deficit angle $\epsilon_f=\beta a -2\pi$ at the triangle. The Euclidean Regge action for gravity restricted to a triangulation of the horizon is given exactly by a sum over faces of the area $A_f$ times this deficit angle,
\begin{equation}
\mathcal{S}_{\text{Regge}}=\frac{1}{8\pi G \hbar}\sum_f A_f(\beta a -2\pi)\;.
\end{equation}
This is exactly the expression in the exponent of the partition function $Z(\beta)$. Moreover this is also the expression of the action that appears in the semiclassical limit of the Spinfoam path integral formulation of Loop Gravity \cite{Barrett:2009mw}. This relation is further explored in \cite{Wieland}, where $Z(\beta)$ is derived from the Spinfoam wedge amplitude $G(\tau)=\langle\Omega|\, \phi \,e^{iK_z a \tau}\,\phi\,|\Omega\rangle$ and the relation with the Gibbons-Hawking path integral is discussed \cite{Gibbons:1976ue}.

\vspace{-1em}
\section{Conclusions}
\vspace{-1em}

We have exploited the fact that the near-horizon geometry of non-extremal black holes is Rindler to derive the Bekenstein-Hawking entropy from Loop Gravity.
We have identified the geometry of a quantum Rindler horizon in Loop Gravity, and shown that it has thermodynamic properties: its temperature agrees with the Unruh temperature of an accelerated observer and its entropy coincides with the Bekenstein-Hawking entropy. The derivation relies only on the dynamics of the Loop Gravity degrees of freedom near the quantum horizon. In particular it provides a quantum version of the notion of horizon entropy density  \cite{Jacobson:2003wv}: each horizon degree of freedom -- a facet dual to a spin-network link puncturing the horizon -- contributes an entropy 
\begin{equation}
s_f=2\pi \gamma j_f\;.
\end{equation}
The analysis does not involve a derivation of the semiclassical black-hole geometry away from the horizon and up to spatial infinity. As a result it extends to all black holes that have a non-degenerate horizon.

The result obtained directly addresses some of the difficulties found in the original Loop Gravity derivation of Black-Hole entropy where the area-ensemble is used \cite{Rovelli:1996dv} and the Immirzi parameter shows up as an ambiguity in the expression of the entropy \cite{Jacobson:2007uj}. Introducing the notion of horizon energy in the quantum theory, we find that the entropy of large black holes is independent from the Immirzi parameter. Quantum gravity corrections to the entropy and the temperature of small black holes are expected to depend on the Immirzi parameter.

\vspace{-1em}
\section{Acknowledgements}
\vspace{-1em}

Thanks to A. Perez, C. Rovelli, H. Haggard, J. Russo, L. Freidel, L. Smolin,  L. Modesto, N. Afshordi, R. Sorkin, T. Jacobson, W. Donnelly, and W. Wieland for many conversations on the problem of black hole entropy. I am especially grateful to C. Rovelli for precious comments and suggestions during the last stage of this work.  I would like to thank also W. Wieland  and J. Hnybida for discussions on the unitary representations of the Lorentz group. Research at Perimeter Institute for Theoretical Physics is supported in part by the Government of Canada through NSERC and by the Province of Ontario through MRI.


\begin{thebibliography}{10}

\bibitem{Bekenstein:1973ur}
J.~D. Bekenstein, 
{\em Phys. Rev.} {\bf D7}  (1973) 2333--2346.

S.~W. Hawking, 
{\em Comm. Math.
  Phys.} {\bf 43} (1975)
199--220.

\bibitem{Rovelli:2004tv}
C.~Rovelli, ``{Quantum gravity},'' Cambridge (2004)

T.~Thiemann, ``{Modern canonical quantum general relativity},'' Cambridge
  Univ. Pr. (2007)

\bibitem{Rovelli:1996dv}
C.~Rovelli, 
{\em Phys. Rev.
  Lett.} {\bf 77} (1996) 3288--3291,
\href{http://arXiv.org/abs/gr-qc/9603063}{{\tt gr-qc/9603063}}.
L.~Smolin, 
 {\em J. Math. Phys.} {\bf 36} (1995) 6417--6455,
\href{http://arXiv.org/abs/gr-qc/9505028}{{\tt gr-qc/9505028}}.

A.~Ashtekar, J.~Baez, A.~Corichi, and K.~Krasnov, 
{\em Phys. Rev. Lett.} {\bf 80} (1998) 904--907,
\href{http://arXiv.org/abs/gr-qc/9710007}{{\tt gr-qc/9710007}}.

J.~Engle, A.~Perez, and K.~Noui, 
{\em Phys. Rev. Lett.} {\bf 105} (2010) 031302,
\href{http://arXiv.org/abs/0905.3168}{{\tt 0905.3168[gr-qc]}}.

\bibitem{Bianchi:2010qd}
E.~Bianchi, 
{\em Class.Quant.Grav.} {\bf 28} (2011) 114006,
\href{http://arXiv.org/abs/1011.5628}{{\tt 1011.5628[gr-qc]}}.

\bibitem{Frodden:2011eb}
E.~Frodden, A.~Ghosh, and A.~Perez,
\href{http://arXiv.org/abs/1110.4055}{{\tt 1110.4055[gr-qc]}}

\bibitem{Engle:2007uq}
J.~Engle, R.~Pereira, and C.~Rovelli, 
  {\em Phys. Rev. Lett.} {\bf 99} (2007) 161301,
\href{http://arXiv.org/abs/0705.2388}{{\tt 0705.2388[gr-qc]}},
et E.~Livine,
  {\em Nucl. Phys.} {\bf B799} (2008) 136--149,
\href{http://arXiv.org/abs/0711.0146}{{\tt 0711.0146[gr-qc]}}.

L.~Freidel and K.~Krasnov, 
{\em Class. Quant. Grav.} {\bf 25} (2008) 125018,
\href{http://arXiv.org/abs/0708.1595}{{\tt 0708.1595[gr-qc]}}.

\bibitem{Unruh:1976db}
W.~Unruh, 
 {\em Phys.Rev.} {\bf D14}
  (1976)
870.

\bibitem{Gibbons:1976ue}
G.~W. Gibbons and S.~W. Hawking, 
  {\em Phys. Rev.} {\bf D15} (1977)
2752--2756.

\bibitem{Wald:1984rg}
R.~M. Wald, ``{General Relativity},'' Chicago U.P. (1984)

\bibitem{Ashtekar:2000sz}
A.~Ashtekar, C.~Beetle, {\em et al.},
  {\em Phys.Rev.Lett.} {\bf 85}
  (2000) 3564--3567,
\href{http://arXiv.org/abs/gr-qc/0006006}{{\tt gr-qc/0006006}}.

\bibitem{Thorne:1986iy}
K.~S. Thorne, R.~H. Price, and D.~A. Macdonald,
  ``{Black Holes: the membrane paradigm},'' Yale (1986)

\bibitem{Ruhl}
Ruhl, {\em The Lorentz group and harmonic analysis},
\newblock Benjamin (1970).
M.~Huszar, 
  {\em Acta Phys. Acad. Sci. Hung.} {\bf 30: No. 3} (1971) 241--51.

\bibitem{Rovelli:2011eq}
C.~Rovelli, 
\href{http://arXiv.org/abs/1102.3660}{{\tt 1102.3660[gr-qc]}}.

\bibitem{DeWitt:1979}
B.~S. DeWitt, 
in General Relativity: an Einstein centenary survey, S. W. Hawking and W. Israel (Eds.), Cambridge
  Univ. Pr. (1979)

\bibitem{Massar:1999wg}
S.~Massar and R.~Parentani, 
  {\em Nucl.Phys.} {\bf B575} (2000) 333--356,
  \href{http://arXiv.org/abs/gr-qc/9903027}{{\tt gr-qc/9903027}}.

\bibitem{Regge:1961px}
T.~Regge, 
{\em Nuovo Cim.} {\bf 19} (1961)
558--571.

\bibitem{Barrett:2009mw}
J.~W. Barrett, R.~Dowdall, W.~J. Fairbairn, F.~Hellmann, and R.~Pereira,
  {\em Class.Quant.Grav.} {\bf 27} (2010) 165009,
\href{http://arXiv.org/abs/0907.2440}{{\tt 0907.2440[gr-qc]}}.

F.~Conrady and L.~Freidel, 
{\em Phys. Rev.} {\bf D78} (2008) 104023,
\href{http://arXiv.org/abs/0809.2280}{{\tt 0809.2280[gr-qc]}}.

E.~Magliaro and C.~Perini, 
\href{http://arXiv.org/abs/1105.0216}{{\tt 1105.0216[gr-qc]}}.

M.~Han and M.~Zhang, 
  \href{http://arXiv.org/abs/1109.0499}{{\tt 1109.0499[gr-qc]}}.

\bibitem{Wieland}
E.~Bianchi and W.~Wieland, to appear.

\bibitem{Jacobson:2003wv}
T.~Jacobson and R.~Parentani, 
{\em Found.Phys.} {\bf 33}
  (2003) 323--348,
\href{http://arXiv.org/abs/gr-qc/0302099}{{\tt gr-qc/0302099}}.

\bibitem{Jacobson:2007uj}
T.~Jacobson, 
{\em Class.Quant.Grav.} {\bf 24} (2007) 4875--4879,
\href{http://arXiv.org/abs/0707.4026}{{\tt 0707.4026[gr-qc]}}.

A.~Ghosh and A.~Perez, 
{\em Phys.Rev.Lett.} {\bf 107} (2011) 241301,
  \href{http://arXiv.org/abs/1107.1320}{{\tt 1107.1320[gr-qc]}}.

\end{thebibliography}

\providecommand{\href}[2]{#2}\begingroup\raggedright\endgroup

\end{document}